\documentclass{article}

\usepackage{PRIMEarxiv}

\usepackage[utf8]{inputenc} % allow utf-8 input
\usepackage[T1]{fontenc}    % use 8-bit T1 fonts
\usepackage{hyperref}       % hyperlinks
\usepackage{url}            % simple URL typesetting
\usepackage{booktabs}       % professional-quality tables
\usepackage{amsfonts}       % blackboard math symbols
\usepackage{nicefrac}       % compact symbols for 1/2, etc.
\usepackage{microtype}      % microtypography
\usepackage{lipsum}
\usepackage{fancyhdr}       % header
\usepackage{graphicx}       % graphics
\graphicspath{{media/}}     % organize your images and other 
\usepackage{algorithm}
\usepackage[noend]{algpseudocode}
\usepackage[T1]{fontenc}
\usepackage[utf8]{inputenc}
\usepackage{xcolor}
\usepackage{mathrsfs}
\usepackage{amssymb}
\usepackage{amsthm}
\usepackage{latexsym,amsmath}
\usepackage{afterpage}
\usepackage{comment}
\usepackage{adjustbox}
\usepackage{caption}
\usepackage{subcaption}
\usepackage{natbib}

\title{Bayesian Copula Directional Dependence for causal inference on gene expression data
}

\author{
  Vasiliki Vamvaka\\
  School of Mathematics and Statistics\\
 University of New South Wales, ACEMS \\
  Sydney, NSW \\
  \texttt{v.vamvaka@student.unsw.edu.au} \\
   \And
  Clara Grazian \\
  	School of Mathematics and Statistics \\
  University of Sydney, ACEMS \\
  	Sydney, NSW\\
  \texttt{clara.grazian@sydney.edu.au} \\
}

\hypersetup{
    colorlinks=true,
    linkcolor=black,
    citecolor=black,
    filecolor=black,
    urlcolor=black,
}

\begin{document}
\maketitle

\begin{abstract}
Modelling and understanding directional gene networks is a major challenge in biology as they play an important role in the architecture and function of genetic systems. Copula Directional Dependence (CDD) can measure the directed connectivity among variables without any strict requirements of distributional and linearity assumptions. Furthermore, copulas can achieve that by isolating the dependence structure of a joint distribution. In this work, a novel extension of the frequentist CDD in the Bayesian setting is introduced. The new method is compared against the frequentist CDD and validated on six gene interactions, three coming from a mouse scRNA-seq dataset and three coming from a bulk epigenome dataset. The results illustrate that the novel proposed Bayesian CDD was able to identify four out of six true interactions with increased robustness compared to the frequentist method. Therefore, the Bayesian CDD can be considered as an  alternative way for modeling the information flow in gene networks.
\end{abstract}

% keywords can be removed

\section{Introduction}
The process of gene expression is observed in each cell of every living organism to determine the cell's functionality and survival. The availability of high-throughput gene expression data in recent years enabled the inference and construction of large-scale Gene Regulatory Networks (GRN). Gene networks are the essential building blocks that control both the expression of proteins and the creation of different types of cells. Gene networks are considered a collection of DNA segments in a cell. These segments interact with each other and with other elements of a cell constructing a network \citep{dubitzky2013encyclopedia}. Modelling and understanding gene interactions is a major challenge in biology as they are considered a “map” for the architecture and function of genetic systems \citep{vijesh2013modeling}. Several methods have been used over the years in order to reconstruct directional GRNs from genomic expression data, such as Boolean network models \citep{shmulevich2002probabilistic, thomas1973boolean, bornholdt2008boolean}, Bayesian networks \citep{zou2005new,friedman2000using} and linear models \citep{chen2005stochastic,deng2005examine}. For a full review the reader can refer to \citet{vijesh2013modeling}.\par
Copulas have become widely used models for analysing multivariate data. The term copula was first introduced by \citet{sklar1959fonctions} and since then they were applied in different areas such as econometrics \citep{huynh2014modeling}, survival analysis \citep{clayton1978model} and medical statistics \citep{lambert2002copula, nikoloulopoulos2008multivariate} among others. The fact that the copula function is able to capture complex forms of dependence between variables makes them suitable to  model gene interactions. Copula-based methods have not been widely used in the literature to infer directional dependence among genes with some exceptions \citep{kim2008copula, kim2009directional}. The authors in \citet{kim2008copula, kim2009directional}, used a method called Copula Directional Dependence (CDD) to infer directional dependences on yeast cell-cycle regulation data. However, these works have several limitations such as the evaluation of the uncertainty in the estimation procedure and the use of parametric copulas, which can lead to estimation biases and model misspecifications. Based on the copula directional dependence method, we propose a novel extension in the Bayesian framework to overcome the aforementioned limitations of the method proposed by \citet{kim2008copula}.\par
The remainder of this article is organized as follows; Section \ref{sec2} provides the background of copula directional dependence, while Section \ref{sec3} illustrates our proposed Bayesian copula directional dependence method. Finally, Section \ref{sec4} presents the results obtained by its application on gene expression data and Section \ref{sec5} concludes the article.

\section{Directional Dependence based on copula regression}
\label{sec2}

In this work, we consider two dimensional copula cases and discuss measures of strength of directional dependence between pairs of random variables. The term directional dependence refers to the likely direction of the influence between two or more variables. The objective of directional dependence is to establish causal relationships among variables. Directional dependence has been studied over the years in different frameworks. The authors in \cite{dodge2000direction} and \cite{muddapur2003directional} discussed the directional dependence in terms of a regression line. By assuming symmetric errors, they concluded that the direction of the dependence between two variables could be determined by looking at their skewness and the Pearson's correlation index. Although the method is straightforward, in real data applications, symmetric errors and linear regression are rarely present. Another limitation of their approach is that the joint behaviour of the variables is not considered. Due to the flexibility of the copula functions to model the dependence among variables, \cite{sungur2005note,sungur2005some} claimed that a copula regression model can be applied to better explain the directional dependence. \cite{sungur2005note} differentiated the terms “direction of dependence” and “directional dependence”, stating that the first is a property of the marginal distributions of the model and the later is derived from their joint behaviour, which is represented by a copula. Copula Directional Dependence (CDD) was proposed as an extension of the directional dependence of  \cite{kim2014copulas} and \cite{kim2014analysis}, as the authors introduced a nonlinear regression based on a logit model \citep{kim2017directional}. The CDD method  is based on a beta regression, following \cite{guolo2014beta}, and the Gaussian Copula Marginal Regression (GCMR) of \cite{masarotto2012gaussian}; this approach estimates the coefficients of the marginal distributions of the regression function as well as it provides inference on the direction of the influence. 

First, the data are appropriately transformed to be uniformly distributed. Let $X_i=(X_{i1},...,X_{id})^{T}$ be a $d$-dimensional random variable, with $i \in (1,...,n)$, be $n$ realizations of a random vector $X$. Pseudo-observations of $X$ are defined as $U_{ij} = F_j(X_{ij})$, where $F_j(\cdot)$ is the marginal distribution of variable $X_{ij}$ for  $i \in (1,...,n)$, $j \in (1,...,d)$. As a result, the variables are defined in the $[0,1]^d$ hypercube. In the case at hand, $(X_1,X_2)$ are measures of gene expression for a pair of genes in the data-set, and they can be transformed into $(U,V) \in [0,1]^2$, where $U=F_1(x_1)$ and $V=F_2(x_2)$.  
Secondly, it is assumed that the transformed variable $V|U=u$ (and similarly $U|V=v$) follows a Beta$(\mu_{v},\kappa_{v})$ distribution parameterized in terms of the mean $\mu_{v}$ and precision parameter $\kappa_{v}$ , as in \cite{ferrari2004beta}, i.e. Beta$(\mu_{v},\kappa_{v})$:
\begin{equation}\label{eq:1}
    f(v; \mu_v,\kappa_v)=\frac{\Gamma(\kappa_v)}{\Gamma(\mu_v\kappa_v)\Gamma((1-\mu_v)\kappa_v)}v^{\mu_v \kappa_v -1}(1-v)^{(1-\mu_v)\kappa_v -1},
\end{equation}

with $0<\mu_v<1$ being the mean parameter, $\kappa_v > 0$ being related to the precision parameter and $\Gamma(\cdot)$ being the gamma function. 
The mean $\mu_v$ can then be related to the conditioning variable $U$ through a logit function
\begin{equation} 
\label{eq:2}
    \text{logit}(\mu_v)  = \frac{\exp(\beta_{0}+\beta_{1}u)}{1
    +\exp(\beta_{0}+\beta_{1}u)},
\end{equation}

and the precision parameter $ \kappa_{v}=1 +exp(\beta_{0}+\beta_{1}u)$.
The parameters of the regression $(\beta_{0},\beta_{1})$ can be estimated with the maximum likelihood method of GCMR \citep{masarotto2012gaussian, ferrari2004beta}. The GCMR separates the marginal components from the dependence structure which, in this case, can be seen as a nuisance parameter. The main advantage of GCMR is that it suggests a natural explanation in terms of copula theory.
By exploiting the probability integral transform, the response $V$ is related to the covariate $U$ and to a standard normal error $\epsilon$ as
\begin{equation}\label{eq:3}
    V_{i}=F^{-1} \{ \Phi(\epsilon);\beta\},  \quad \mbox{for} \quad   i=1,...,n
\end{equation}

where $F(\cdot,\beta)$ and $\Phi(\cdot)$ are the cumulative distributions of $V|U=u$ and $\epsilon$ respectively. Finally, measures of strength of directional dependence between pairs of random variables can be derived through a copula representation.\par
According to Sklar's theorem \citep{sklar1959fonctions}, any joint cumulative distribution function can be written in terms of a copula function $C$. Precisely, for two random variables $(X_1,X_2)$, the joint distribution $F_{X_{1},X_{2}}(X_1,X_2)$ can be written as
\begin{equation}
\label{eq:4}
F_{X_{1},X_{2}}(x_1,x_2)=C(F_1(x_1),F_2(x_2)) = C(u,v)
\end{equation}

where $F_1, F_2$ are the marginal distributions of $X_1$ and $X_2$, respectively. Let $C(u,v)$ denote a copula of two uniformly distributed random variables $(U,V)$ and $C_{V|U}$ denote the conditional distribution of $V|U=u$, defined as 
\begin{equation}
\label{eq:5}
C_{V|U} \equiv C_u(v) \equiv  P(V \leq v | U =u) = \frac{\partial{C(u,v)}}{\partial{u}}.
\end{equation}

Then, the mean of $V$ given $U$, denoted as $r_{V|U=u}$, can be expressed in terms of the copula function as
\begin{equation} 
\label{eq:6}
r_{V|U=u} \equiv E(V | U = u) = 1- \int_{0}^{1} C_{u}(v)dv.
\end{equation}

\cite{sungur2005note} also introduced a measure to infer the strength of directional dependence between two variables based on their copula regression
\begin{equation}
\label{eq:7}
\rho_{U\rightarrow V}^{2}=\frac{Var(r_{V|U=u})}{Var(V)}=\frac{E[(r_{V|U=u})-1/2)^2]}{1/12} = 12E[(r_{V|U}(u))^2]-3,
\end{equation}

i.e. $\rho_{U\rightarrow V}^{2}$ can be interpreted as the proportion of total variation of $V$ that can be explained by the copula regression of $V$ on $U$. If $U$ and $V$ are independent then, $C(u,v) = uv$ and thus $r_{V|U=u}$ and $r_{U|V=v}$ are both equal to $0.5$. It is also easy to notice that the copula directional dependence measure of Equation \eqref{eq:7} is a version of Spearman's $\rho$ correlation coefficient, which can be expressed in copula terms as
\begin{equation}\label{eq:8}
\rho_{C} = 12 \int_{0}^{1} \int_{0}^{1} C(u,v)dudv -3= 12 \int_{0}^{1} \int_{0}^{1} (C(u,v) - uv)dudv.
\end{equation}

By combining Equations \eqref{eq:2} and \eqref{eq:7}, the measure of the directional dependence takes the form
\begin{equation} 
\label{eq:9}
    \rho_{U\rightarrow V}^{2}=\frac{Var(E(V|=u))}{Var(V)}=12Var(\mu_{v})=12\sigma_{v}^{2}
\end{equation}

where $\sigma_{v}^{2} = Var(\mu_{v})$. Similar formulas can be derived to construct the directional dependence measure for the other direction, i.e. from $V$ to $U$ based on the variable $U | V=v$. 
Finally, the resulting CDD estimates, $\rho_{V\rightarrow U}^{2}$ and $\rho_{U\rightarrow V}^{2}$ can be compared and used to identify the strongest direction, i.e. the highest value between $\rho_{V\rightarrow U}^{2}$ and $\rho_{U\rightarrow V}^{2}$ indicates the direction of influence among the two variables. \par
Test of the type $H_0: \; \rho^2_{V \rightarrow U} \leq \rho^2_{U \rightarrow V}$ against $H_0: \; \rho^2_{V \rightarrow U} > \rho^2_{U \rightarrow V}$ can be carried out by comparing the confidence intervals of the differecne $\Delta \rho_{U,V}^2$ of the two statistics $\rho^2_{V \rightarrow U}$ and $\rho^2_{U \rightarrow V} $. However, confidence intervals can only be obtained through bootstrap techniques, which are known to underestimate the uncertainty of the estimation procedure in case of copula measures; see, for example, \cite{grazian2017approximate} for more details.

\section{Bayesian Copula Directional Dependence}
\label{sec3}

We propose to extend the copula directional dependence model of \cite{kim2017directional}, to a Bayesian framework: in this way, uncertainty is better evaluated and more interpretable results can be achieved, since credible intervals based on the posterior distribution of $\rho^2_{V \rightarrow U}$ and $\rho^2_{U \rightarrow V}$, approximated via a Markov Chain Monte Carlo (MCMC) algorithm, are easily obtained.\par
As described in Section \ref{sec2}, it is assumed that $V|U=u$ follows a Beta$(\mu_v, \kappa_v)$ distribution with mean $\mu_v$ as seen in Equation \eqref{eq:7}. Similarly, $U|V=v$ follows a Beta$(\mu_u, \kappa_u)$. To derive the posterior distribution of the coefficients of the beta regression, prior distributions need to be assigned. The choice of priors should be made according to the knowledge of the problem and how much information needs to be included. For the purpose of this work, weakly informative priors are introduced for the coefficients of the logit model described in Equation \eqref{eq:1} . More precisely,
\begin{equation} \label{eq:10}
\begin{aligned}
   \beta^{(v)}_{0} \sim \mathcal{N}(0,\sigma^2_0), \qquad \beta^{(v)}_{1} \sim \mathcal{N}(0,\sigma^2_1),
    \end{aligned}
\end{equation}
where $\sigma_0$ and $\sigma_1$ are both chosen to be equal to $10$. Here $\beta^{(v)}_{\ell}$ indicates the coefficient of the beta regression of $V$ given $U=u$, for $\ell = 0,1$, and, similarly, $\beta^{(u)}_{\ell}$ indicates the coefficients of the beta regression of $U$ given $V=v$. 

Moreover, two cases for the precision parameter $\kappa$ were investigated. In the first case, following the frequentist CDD theory from Section \ref{sec2}, the precision parameter was fixed as
\begin{equation} 
\label{eq:11}
\kappa_{v} = 1+\exp (\beta^{(v)}_{0}+\beta^{(v)}_{1}u).
\end{equation}
An advantage of fixing $\kappa_{v}$ is that the computation burden of the algorithm is lower, as there are two parameters in the model that need to be estimated; $\beta^{(v)}_{0}$ and $\beta^{(v)}_{1}$. Alternatively, $\kappa_v$ can be defined independently from $\beta^{(v)}_{0}$ and $\beta^{(v)}_{1}$ and a Gamma prior distribution can be assigned, i.e. $\kappa_v \sim \Gamma (a,b)$; here $a=b=1$. While the computational burden increases, this version of the model was noticed to introduce more variability in the simulations and helped to stabilize the estimation process of $\kappa_v$.

\begin{algorithm}
\caption{Parametric Bayesian Copula Directional Dependence}\label{alg:PBCDD}
\begin{algorithmic}[1]

\State Transform $(X_1,X_2)$ genes to $(U,V) \in [0,1]^2$.
\vspace{3mm}
\State Initialize $\beta^{(v)}_{0}, \beta^{(v)}_{1}, \kappa_v$ and $\beta^{(u)}_{0}, \beta^{(u)}_{1}, \kappa_u$.
\vspace{3mm}
\State Set:
\begin{align*}
    \mu_{v}  &=\frac{exp(\beta^{(v)}_{0}+\beta^{(v)}_{1}v)}{1
    +exp(\beta^{(v)}_{0}+\beta^{(v)}_{1}v)} & \mu_{u}  &=\frac{exp(\beta^{(u)}_{0}+\beta^{(u)}_{1}u)}{1
    +exp(\beta^{(u)}_{0}+\beta^{(u)}_{1}u)}
\end{align*}
\State For $t=2,\ldots,T$:\par
Propose new values \par
\begin{align*}
\Big (\beta^{*(v)}_{0}, \beta^{*(v)}_{1} \Big) &\sim \mathcal{N}\Big (\beta^{(t-1)(v)}_{0}, \beta^{(t-1)(v)}_{1},\Sigma \Big) \\
\Big (\beta^{*(u)}_{0}, \beta^{*(u)}_{1} \Big) &\sim \mathcal{N}\Big (\beta^{(t-1)(u)}_{0}, \beta^{(t-1)(u)}_{1},\Sigma \Big) \\
\kappa^{*}_{v}&\sim \mathcal{N}\Big (\kappa^{(t-1)}_{v},\Sigma \Big)\\
\kappa^{*}_{u}&\sim \mathcal{N}\Big (\kappa^{(t-1)}_{u},\Sigma \Big)
\end{align*}

and accept proposed values $\Big (\beta^{*(v)}_{0}, \beta^{*(v)}_{1} \Big)$ and $\Big (\beta^{*(u)}_{0}, \beta^{*(u)}_{1} \Big)$ with probability: \par
\begin{align*}
    \alpha_v &= \min\left\{1, \frac{\pi\Big ( \beta^{*(v)}_0,\beta^{*(v)}_1,\kappa^*_v|u,v \Big) q(\kappa_v^{(t-1)})}{\pi\Big ( \beta^{(t-1)(v)}_0,\beta^{(t-1)(v)}_1,\kappa^{(t-1)}_v|u,v \Big) q(\kappa_v^{*})} \right\} \\
    \alpha_u &= \min\left\{1, \frac{\pi\Big ( \beta^{*(u)}_0,\beta^{*(u)}_1,\kappa^*_u|u,v \Big) q(\kappa_u^{(t-1)})}{\pi\Big ( \beta^{(t-1)(v)}_0,\beta^{(t-1)(v)}_1,\kappa^{(t-1)}_v|u,v \Big) q(\kappa_u^{*})} \right\} 
\end{align*}
\vspace{3mm}
\State \textbf{Output:} a sample of values for the coefficients $\Big(\beta^{(v)}_{0}, \beta^{(v)}_{1},\kappa_v \Big)$ and $\Big(\beta^{(u)}_{0}, \beta^{(u)}_{1},\kappa_u \Big)$ approximately from their posterior distribution.
\vspace{3mm}
\State Use posterior samples of \textbf{step 5} to calculate:
\begin{align*}
\rho^2_{U\rightarrow V}&=12Var(\mu_{v})& \rho^2_{V\rightarrow U}&=12Var(\mu_{u})\end{align*}
\State \textbf{if} {$\Pr\left(\rho^2_{U\rightarrow V}>\rho^2_{V\rightarrow U} | u,v \right) > 0.5$}
     \textbf{then} the identified direction is from $U$ to $V$
\end{algorithmic}
\end{algorithm}
The posterior distribution for the parameters of the beta regression are given by
\begin{equation} 
\label{eq:12}
\pi\Big ( \beta^{(v)}_0,\beta^{(v)}_1,\kappa_v|u,v \Big) \propto 
\pi\Big(\beta^{(v)}_0\Big)\pi\Big(\beta^{(v)}_1\Big)\pi(\kappa_v)\mathcal{L} \Big (u,v|\beta^{(v)}_0,\beta^{(v)}_1,\kappa_v \Big )
\end{equation}
where $\pi(\beta^{(v)}_0)$, $\pi(\beta^{(v)}_1)$, $\pi(\kappa_v)$ are the prior distributions for $\beta_0^{(v)},\beta^{(v)}_1,\kappa_v$ as defined in Equation \eqref{eq:10} and $\mathcal{L} \Big (u,v|(\beta_0,\beta_1,\kappa) \Big )$ is the likelihood function of the transformed gene data, which comes from a Beta distribution as in Equation \eqref{eq:1}. Similarly expressions can be obtained for the parameters of the model for $U | V=v$. 

The posterior distribution of Equation \eqref{eq:12} can be approximated via MCMC. In particular, we use random walk Metropolis-Hastings, with normal proposal distributions, as described in Algorithm \ref{alg:PBCDD}.
After posterior samples are obtained for the parameter of the beta regression model for $V|U=u$ and $U|V=v$, this posterior distribution induces a posterior distribution on $\rho^2_{U \rightarrow V}$ and $\rho^2_{V \rightarrow U}$, and credible intervals can be derived. More specific, it is possible to approximate the $\Pr\left(\rho^2_{U\rightarrow V}>\rho^2_{V\rightarrow U} | u,v \right)$ and identify the most likely direction of influence depending on some threshold, for example, 0.5.

\section{Application to gene expression data}\label{sec4}
The two data sets under investigation arise from two different sequencing techniques, single-cell sequencing and bulk RNA sequencing. The first data set provides information about a mouse single-cell RNA-seq (scRNA-seq) and the second one is a bulk epigenome dataset, which comprises different regulatory data types from eight different genetic lines of \textit{Drosophila} embryos.

The mouse scRNA-seq data set, contains single-cell measurements of cells in the lungs called  alveolar type 2 (AT-2). The AT-2 cells, found in the respiratory area of mammals, are responsible for lowering the surface tension in the lungs which is caused during gas exchange. The scRNA-seq mouse data set consists of observations of the thyroid transcription factor Nkx2.1 with the pulmonary surfactant proteins Sftpa1, Sftpb and Sftpc respectively. The pulmonary surfactant proteins are regulated by the NK2 homeobox 1, also called thyroid transcription factor 1 (Nkx2.1), which attaches to DNA and manages their activation \citep{cao2010epigenetic}.

The epigenome is composed of chemical modifications to DNA and histone proteins, which regulate the formation of chromatin and the function of the genes \citep{bernstein_meissner_lander_2007}. The mechanisms of epigenetics are the ones responsible for the changes in the way RNA interacts with DNA and influences the expression of the genes. The bulk epigenome data set contains several measurements from 8 genetic lines of \textit{Drosophila} embryos at different embryonic stages. The features of the data set include whole-genome profiles of RNA-seq, ATAC-seq, H3K4me3 and H3K27ac. Histone H3 is one of the main proteins forming chromatin. It consists of five lysines, K4, K9, K27 and K79 \citep{sims2003histone}. Histone mark enrichments H3K4me3 and H3K27ac are both involved in transcriptionally active genes. Furthermore, H3K4me3 methylation is generally linked with the activation of chromatin and with having the instructive role in the gene activation \citep{santos2002active}. H3K27ac is defined as an active enhancer mark, which bounds proteins and increases the probability that a gene will be activated. Assay for Transposase-Accessible Chromatin using sequencing (ATAC-seq) is used in epigenomic analysis for inference of chromatin accessibility \citep{buenrostro2013transposition}.

%% frequentist estimates
\begin{table}[ht]
\caption{Estimates of the copula directional dependences. $\Delta \rho^{2}$ denotes the difference $\Delta\rho^{2}=\rho_{U\rightarrow V}^{2}- \rho_{V\rightarrow U}^{2}$. $LB(\Delta \rho^{2})$ and  $UB(\Delta \rho^{2})$ represent the lower and upper bound of the $95\%$ confidence interval for the difference, $\Delta \rho^{2}$, respectively. The entries in bold indicate the stronger direction.}
\centering
\begin{tabular}{l l c c c c c}
\toprule
Gene U &Gene V & $\rho_{U\rightarrow V}^{2}$ & $\rho_{V\rightarrow U}^{2}$ & $\Delta \rho_{U,V}^{2}$ & $LB(\Delta \rho_{U,V}^{2})$ & $UB(\Delta \rho_{U,V}^{2})$\\
\midrule
Nkx2.1 & Sftpa1  & $\textbf{ 0.0004167}$ &$0.0000019$&$0.0004148$&$0.0002928$& $0.0004700$ \\
Nkx2.1 & Sftpb &  $0.0573268$  & $\textbf{0.0686556}$ & $-0.0113288$ & $0.0106055$& $0.0115282$\\
Nkx2.1 & Sftpc  & $\textbf{0.0003475}$&  $0.0000003$&$0.0003472$& $0.0004257$ &$0.0006244$ \\
ATAC-seq & H3K27ac  &$0.0112227$ & $\textbf{0.0142323}$ &$-0.0030096$& $0.0029179$&$ 0.0030393$\\
ATAC-seq & RNA-seq  & $0.0497491$& $\textbf{0.0498537}$& $-0.0001047$&$-0.0001054$& $0.0002875$\\
H3K4me3 & RNA-seq &$\textbf{ 0.0234852}$  &$0.0211191$& $0.0023660$& $0.0020347$&$0.0024177$\\
\bottomrule
\label{table:1}
\end{tabular}
\end{table}

%%% Results----------------------------------------------------
The CDD method of \cite{kim2017directional} was able to identify four out of six directionalities correctly as observed in Table \ref{table:1}. 
When applied to the scRNA-seq data set, the algorithm was able to identify correctly the information flow from Nkx2.1 to the surfactant proteins Sftpa1 and Sftpc (Nkx2.1 $\rightarrow$ Sftpa1, Nkx2.1 $\rightarrow$ Sftpc), but wrongly describes the relationship between Nkx2.1 and Sftpb. When applied to the bulk sequencing dataset, it was able to estimate the correct directional dependence of two gene pairs; H3K27ac $\rightarrow$ ATAC-seq and H3K4me3$\rightarrow$ RNA-seq. The result for the pair RNA-seq $\rightarrow$ ATAC-seq is not statistically significant, as the lower bound of the $95\%$ confidence interval of the difference was found to be negative. Furthermore, as Table \ref{table:1} depicts, the confidence intervals in all gene pairs are highly underestimated and do not provide sufficient coverage for reliable causal inference.  This drawback can be due to the fact that potential issues are raised when ranking measures, such as $\Delta \rho_{U,V}^{2}$, are bootstraped. Furthermore, gene expressions present extreme values in their distributions and bootstrap methods may underestimate the variability of the observations.

%% Bayesian CDD estimates

\begin{table}[ht]
\caption{Estimates of the copula directional dependences of the parametric Bayesian CDD algorithm. $\Delta \rho_{U,V}^2$ denotes the difference $\Delta \rho^{2}_{U,V}=\rho^{2}_{U\rightarrow V}-\rho^{2}_{V\rightarrow U}$. The mean and $95\%$ credible intervals are given for each estimate. The entries in bold indicate the stronger direction.}
\centering
\begin{tabular}{l l c c c }
\toprule
Gene U & Gene V & $\boldsymbol{\rho^{2}_{U\rightarrow V}}$ &$\boldsymbol{\rho^{2}_{V\rightarrow U}}$&$\boldsymbol{\Delta \rho^{2}_{U,V}}$\\
\phantom{abc} & \phantom{abc} & $95\%CI$ &$95\%CI$&$95\%CI$\\
 
\midrule

Nkx2.1 & Sftpa1  &  $ 0.0073155$ &$\textbf{ 0.0149348}$& $-0.0088193$\\
\phantom{abc}&\phantom{abc}&\footnotesize{$(0.0000408,0.0290095)$}&
\footnotesize{$(0.0000767, 0.051982)$}&\footnotesize{$(-0.0323515,0.0145573)$}\\

Nkx2.1 & Sftpb &  $ 0.0070168$ & $\textbf{0.0108458}$ & $-0.0046731$\\
\phantom{abc}&\phantom{abc}&\footnotesize{$(0.0000258,0.0273997)$}&
\footnotesize{$(0.0000752,0.0402749) $}&\footnotesize{$(-0.0281410,0.0169384) $}\\

Nkx2.1 & Sftpc  & $ \textbf{ 0.070041}$& $0.014552$& $ 0.0568564$ \\
\phantom{abc}&\phantom{abc}&\footnotesize{$(0.0178034,0.140973) $}&
\footnotesize{$(0.0000832,0.0498539) $}&\footnotesize{$(0.0095509,0.1049905)$}\\

ATAC-seq & H3K27ac  &$0.10595$ & $\textbf{ 0.1585976}$&$-0.0527481$\\
\phantom{abc}&\phantom{abc}&\footnotesize{$ (0.0861014,0.1273347)$}&
\footnotesize{$(0.1385523,0.1792747) $}&\footnotesize{$(-0.0707538,0.0363282)$}\\

ATAC-seq & RNA-seq  & $  0.0999059$& $ \textbf{0.146957}$& $-0.047137$\\
\phantom{abc}&\phantom{abc}&\footnotesize{$(0.0621321,0.1422733) $}&
\footnotesize{$(0.1043268, 0.1897011)$}&\footnotesize{$(-0.0794135, -0.0159033)$}\\

H3K4me3 & RNA-seq &$ \textbf{0.07501433}$ & $ 0.0346218$&$  0.0432002$\\
\phantom{abc}&\phantom{abc}&\footnotesize{$(0.0297109,0.1325998) $}&
\footnotesize{$(0.0048508, 0.0782323) $}&\footnotesize{$( 0.0061894, 0.0782323)$}\\
\bottomrule
\label{table:3}
\end{tabular}
\end{table}

\begin{figure}[H]
  \begin{subfigure}[b]{0.5\textwidth}
    \includegraphics[width=\textwidth]{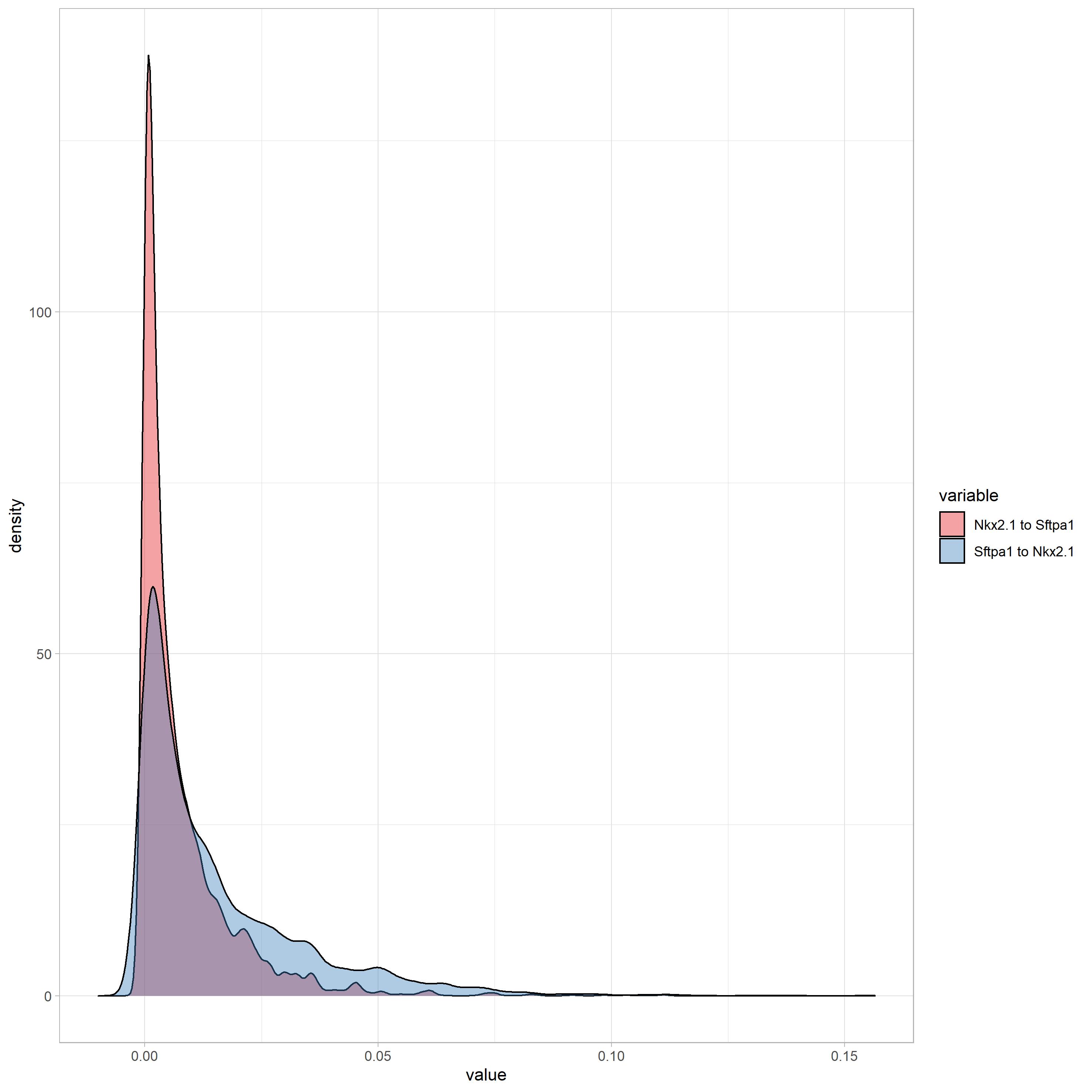}
    \caption{Nkx2.1 and Sftpa1}
    \label{figpp1}
  \end{subfigure}
  \begin{subfigure}[b]{0.5\textwidth}
    \includegraphics[width=\textwidth]{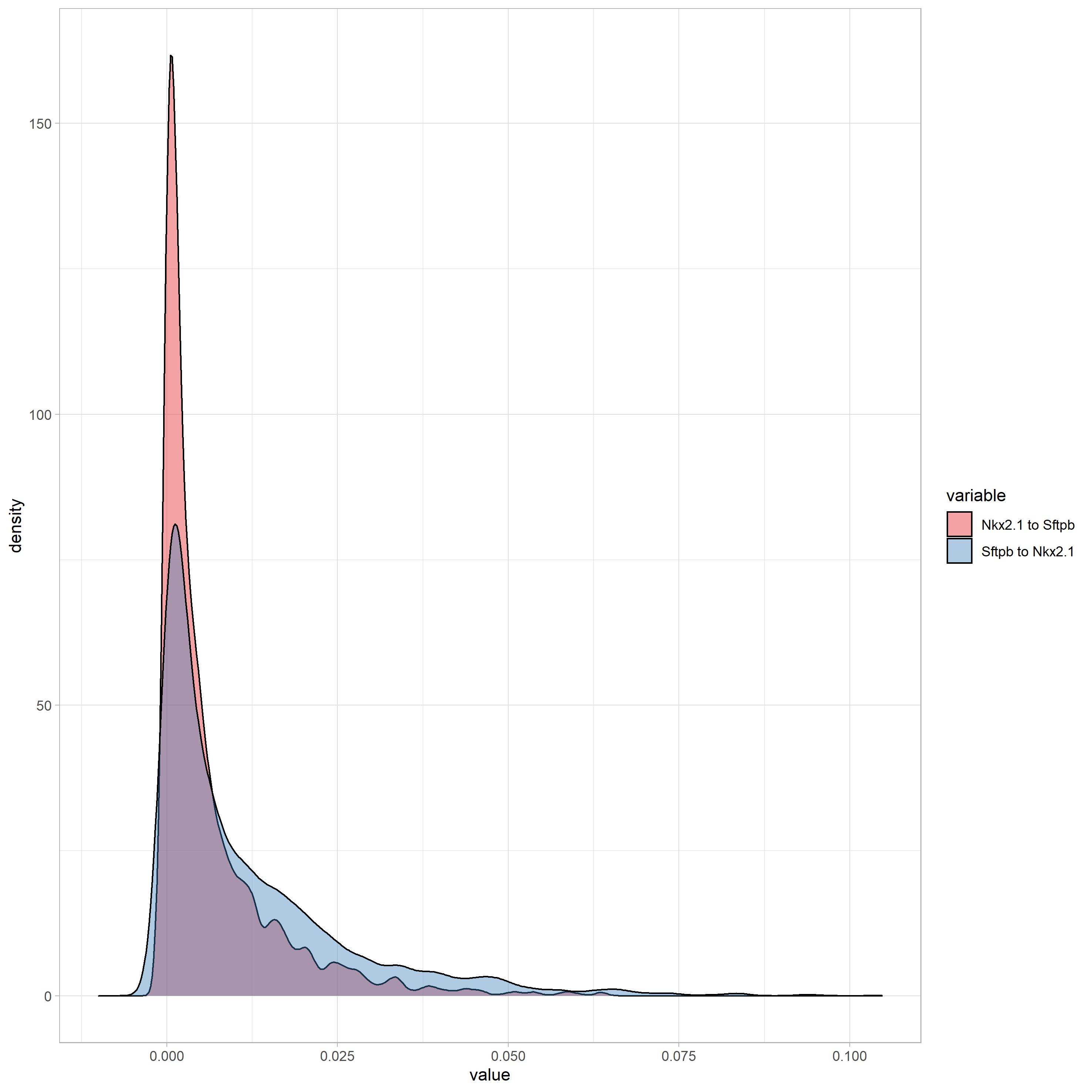}
    \caption{Nkx2.1 and Sftpb}
    \label{figpp2}
  \end{subfigure}
    \end{figure}
  \begin{figure}\ContinuedFloat
    \begin{subfigure}[b]{0.5\textwidth}
    \includegraphics[width=\textwidth]{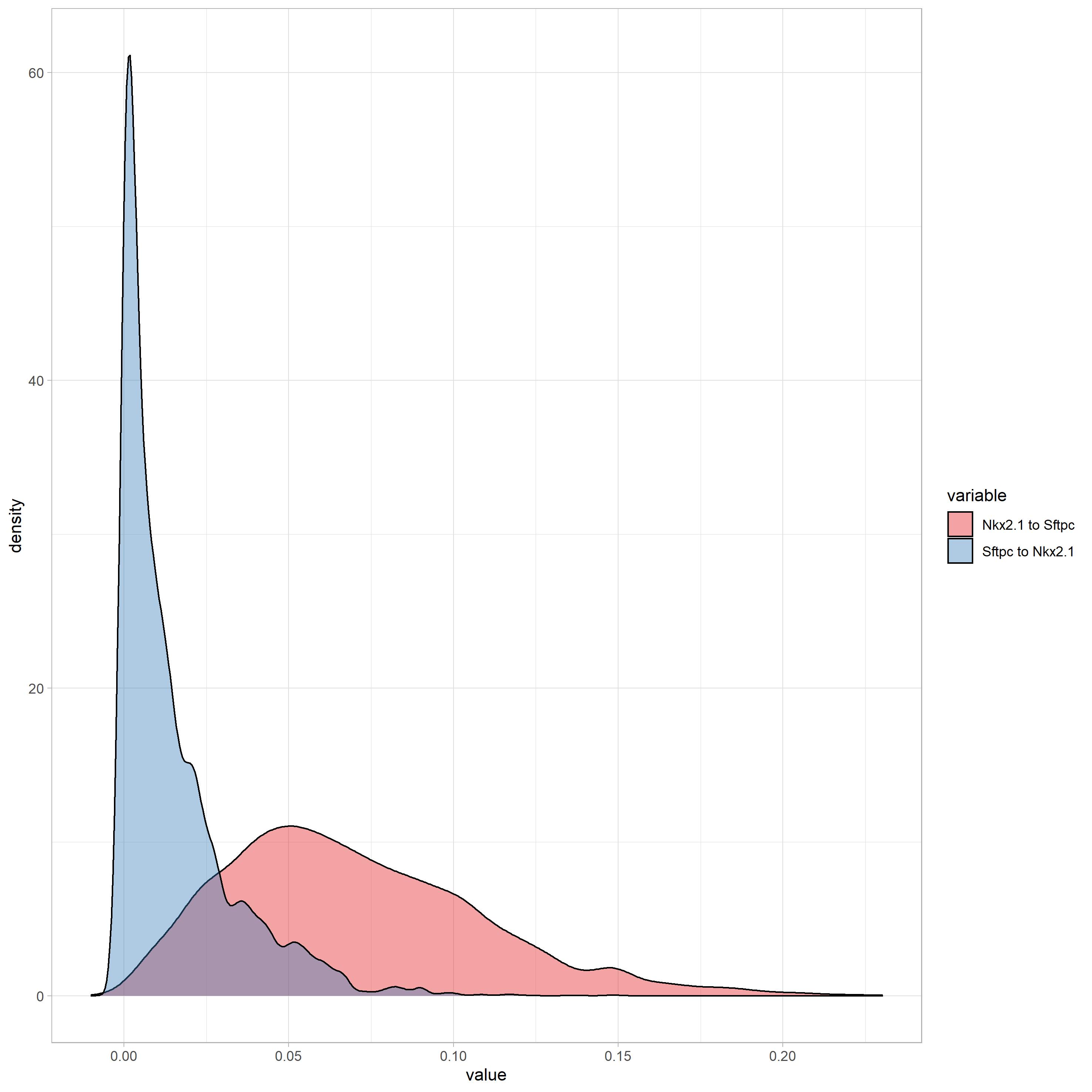}
    \caption{Nkx2.1 and Sftpc}
    \label{figpp3}
  \end{subfigure}
  \begin{subfigure}[b]{0.5\textwidth}
    \includegraphics[width=\textwidth]{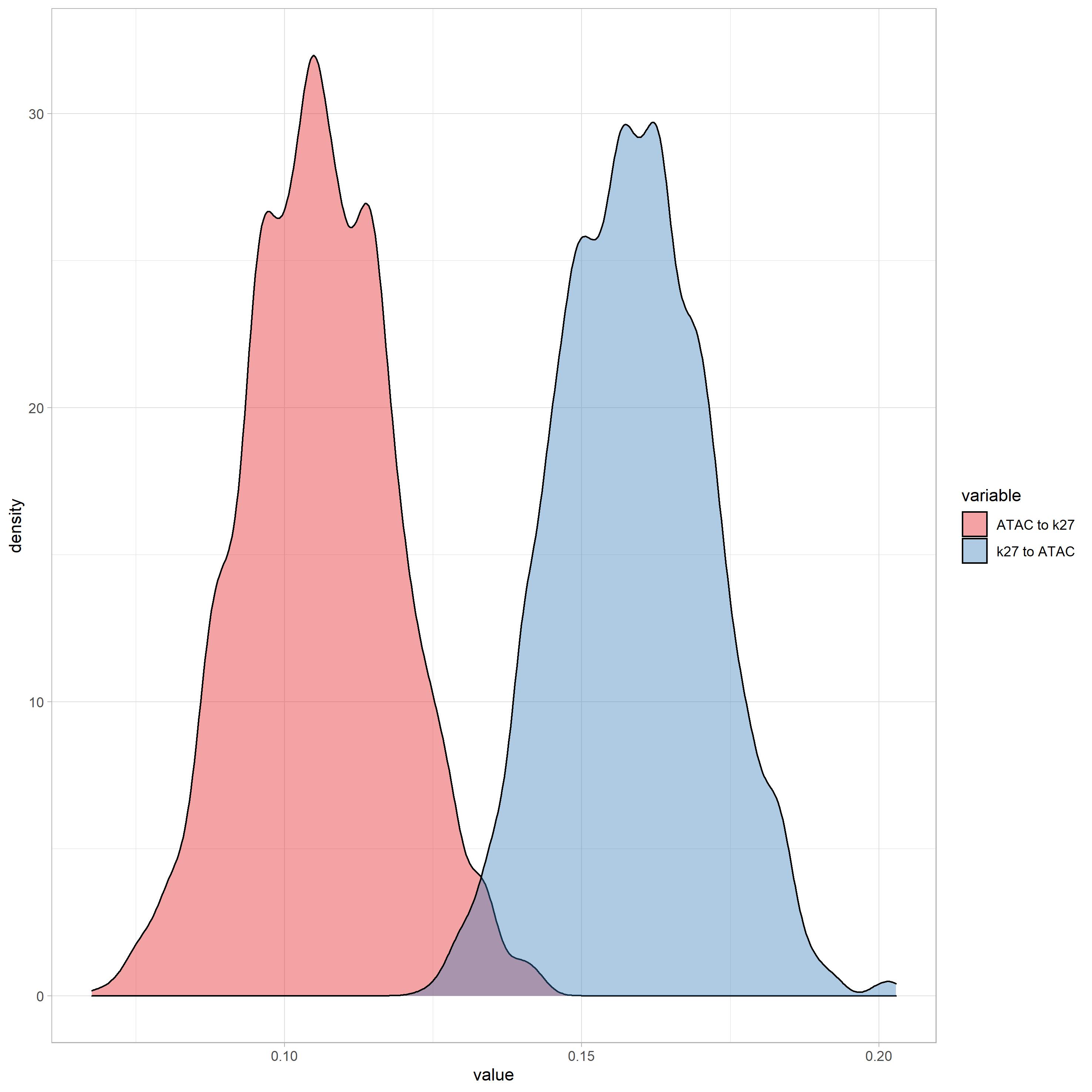}
    \caption{ATAC and k27}
    \label{figpp4}
  \end{subfigure}
  %\end{figure}
 %\begin{figure}\ContinuedFloat
    \begin{subfigure}[b]{0.5\textwidth}
    \includegraphics[width=\textwidth]{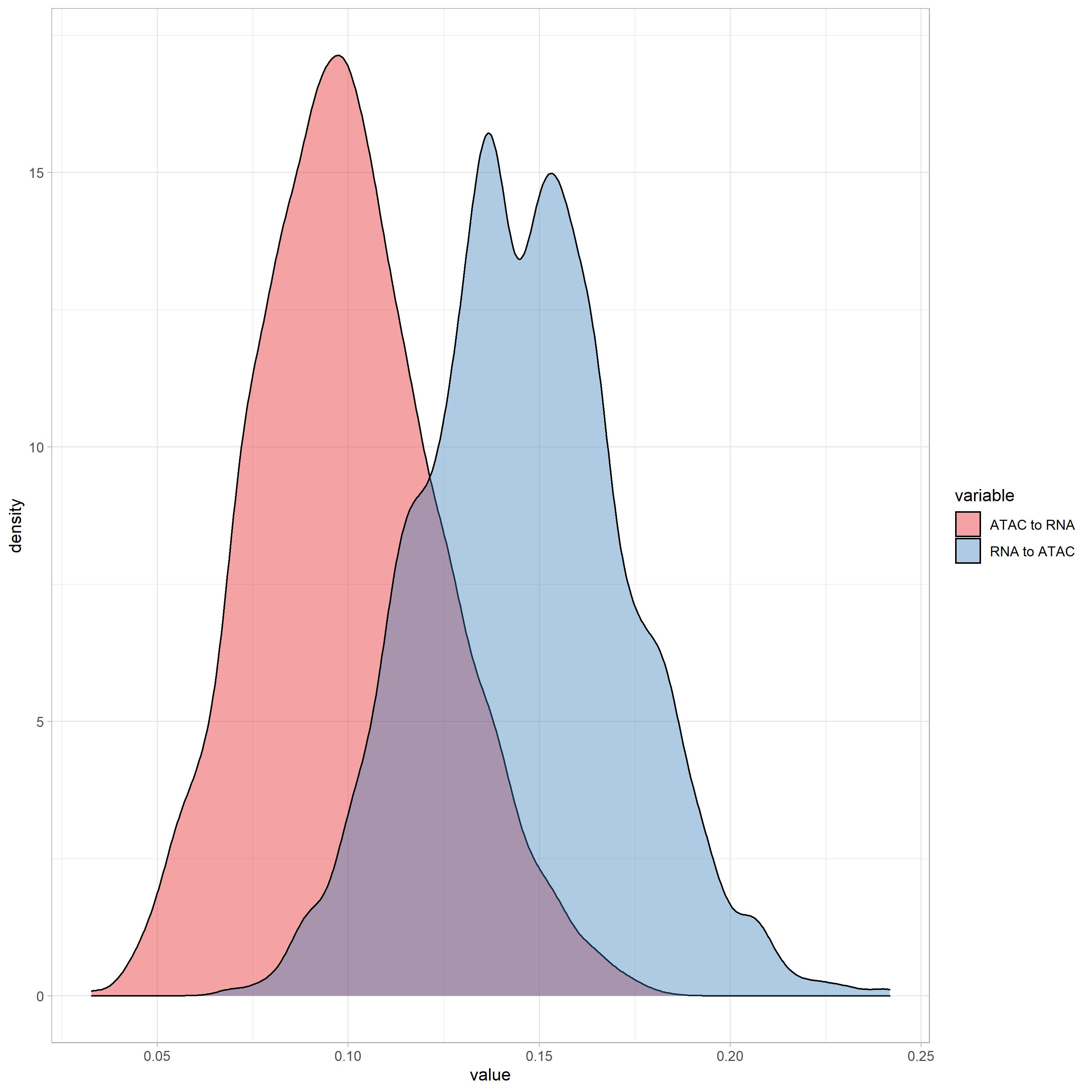}
    \caption{ATAC and RNA}
    \label{figpp5}
  \end{subfigure}
  \begin{subfigure}[b]{0.5\textwidth}
    \includegraphics[width=\textwidth]{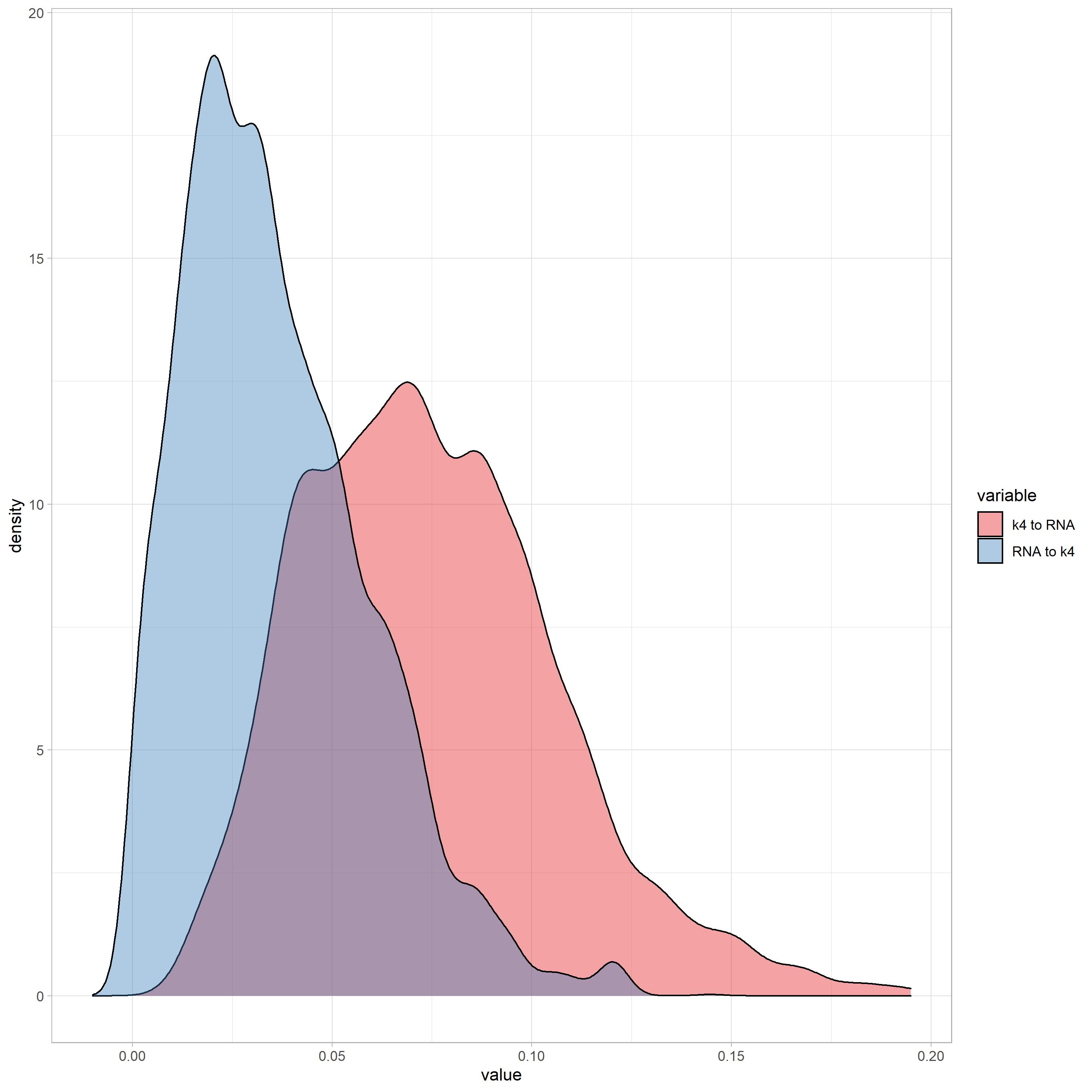}
    \caption{k4 and RNA}
    \label{figpp6}
  \end{subfigure}
  \caption{Posterior densities of the Bayesian parametric CDD method for each pair of genes. The red densities represent the $\rho_{U\rightarrow V}^2$ and the blue the $\rho_{V\rightarrow U}^2$ posterior CDD samples.} 
  \label{figpp}
\end{figure}
\newpage
The parametric Bayesian CDD method was able to capture four out of six correct directionalities. The correct causal relationships that were retrieved are Nkx2.1 $\rightarrow$ Sftpc, H3K27ac $\rightarrow$ ATAC-seq, RNA-seq $\rightarrow$ ATAC-seq and H3K4me3 $\rightarrow$ RNA-seq. Table \ref{table:3} summarizes the mean values of the estimates of the posterior densities, $\rho_{U\rightarrow V}^2$, $\rho_{V\rightarrow U}^2$, their difference $\Delta \rho_{U,V}^2$ and their ($5\%,95\%$) credible intervals, while Table \ref{table:4} depicts the probabilities suggesting each direction of influence as a percentage. Figure \ref{figpp} illustrates the plots for the posterior density distributions for the Bayesian CDD estimates. As presented in Figures \ref{figpp1} and \ref{figpp2}, the posterior directional dependence densities for the pairs Nkx2.1-Sftpa1 and Nkx2.1-Sftpb are highly overlapped, hence the direction of influence is decided only by a few samples. On the other hand, as seen in Figure \ref{figpp3}, the densities for the pair Nkx2.1-Sftpc are more distinct, with $79\%$ of the posterior samples coming from the correct direction. Overall, the algorithm performed with higher accuracy in the bulk epigenome dataset compared to the scRNA-seq data, as it was able to identify all three directional dependences correctly with a probability of more than $87\%$ in each pair. Furthermore, for genes belonging to the bulk sequencing, the percentage of the frequencies indicating the correct and stronger direction are very high, resulting in distinct and clear posterior distributions as shown in Figures \ref{figpp4} to \ref{figpp6}.\par
%%% Percentages from each direction-----------------------------------------------

\begin{table}[ht]
\caption{Percentage of posterior samples suggesting the direction of influence for each pair of genes for the parametric Bayesian CDD algorithm.}
\centering
\begin{tabular}{l l c c }
\toprule
Gene U & Gene V & Samples $\rho^{2}_{U \rightarrow V}>\rho^{2}_{V \rightarrow U}$  &Samples $\rho^{2}_{U \rightarrow V}<\rho^{2}_{V \rightarrow U}$  \\[0.3cm]
\midrule
Nkx2.1 & Sftpa1 &  $41.2\%$  & $58.8\%$\\
Nkx2.1 & Sftpb &  $45.6\%$  & $54.4\%$\\
Nkx2.1 & Sftpc &  $79\%$  & $21\%$\\
ATAC-seq &H3K27ac &  $0.1\%$  & $99.9\%$\\
ATAC-seq & RNA-seq &  $8.1\%$  & $91.9\%$\\
H3K4me3 &RNA-seq &  $87.5\%$  & $12.5\%$\\
\bottomrule
\label{table:4}
\end{tabular}
\end{table}

The biological interpretation of the inferred directional dependendces are as follows.
The gene Nkx2.1 is the transcription factor which regulates proteins Sftpa1, Sftpb and Sftpc. The association between Nkx2.1 and Sftpc was described correctly by both the frequentist and the Bayesian CDD. This  result can be explained by the fact that during our exploratory analysis, we noticed that the Sftpc gene was the most highly expressed among the three surfactant proteins, so the results were significant only for this pair. This result is also highlighted visually in the posterior CDD distributions in the Bayesian method in Figure \ref{figpp3}.
The interaction H3K27ac $\rightarrow$ ATAC-seq, implies that the histone mark k27 is enriched at regions that bind proteins. These proteins are the ones that open up chromatin ATAC-seq, the gene that regulates chromatin accessibility. The influence of RNA-seq $\rightarrow$ ATAC-seq was highlighted only by the proposed Bayesian CDD method. The result suggests that variation in RNA-seq is buffered compared to ATAC-seq. A cause for this behaviour, is that RNA-seq is the functional component that gets translated into proteins, hence there exist a mechanism in place to ensure it is not too sensitive to changes in chromatin. Finally, the direction of influence H3K4me3 $\rightarrow$ RNA-seq agrees with the biological interaction as mark enrichment H3K4me3 is always present where gene expression exists, measured from RNA-seq. A comparison of the performance of the two methods is showcased in Table \ref{table:5}.\par

\begin{table}[h]
\caption{Comparison of performance between the frequentist CDD and the Bayesian CDD methods.The symbol O indicates that the interaction is captured, whereas the symbol X indicates that the gene interaction was not captured by the respective method. }
\centering
\begin{tabular}{l c c }
\toprule
True gene interaction & Frequentist CDD & Bayesian CDD  \\[0.3cm]
\midrule
Nkx2.1 $\rightarrow$ Sftpa1 &  O & X\\
Nkx2.1 $\rightarrow$ Sftpb &  X & X\\
Nkx2.1 $\rightarrow$ Sftpc &  O  & O\\
H3K27ac $\rightarrow$ ATAC-seq&  O  & O\\
RNA-seq $\rightarrow$ ATAC-seq&  X  & O\\
H3K4me3 $\rightarrow$ RNA-seq &  O  & O\\
\bottomrule
\label{table:5}
\end{tabular}
\end{table}

%%%%%%%---- CONCLUSION-------------------------------------------------------------------------------------------
\section{Conclusion}\label{sec5}
In this work, causal inference among gene pairs of both single-cell and bulk sequencing data was investigated. Copula based methods were chosen to be practised, since they are able to isolate the marginal effect of the distributions from the joint behaviour. That way, one is able to consider the direction of influence through a functional of the copula structure, where the direction of the dependence lies. Frequentist Copula Directional Dependence introduced by \cite{kim2017directional} and our novel proposed method of Bayesian Copula Directional Dependence, were applied on two gene expression data sets. Their performance was evaluated in terms of their ability to produce accurate and informative results according to the true biological interaction between the genes.\par
A common observation for both methods is that no particular copula function was applied to the data, since there was no prior evidence that a specific family is appropriate. This is extremely useful for implementations where only the directionality rather than the complete dependence structure is of interest. The CDD measures infer connectivity, so the relative strengths of the directional dependences can be compared between gene pairs, based on nonlinear copula regression models. Although the frequentist CDD retrieved four out of six interactions, the results were not promising, as the directional dependence estimates and their confidence intervals were highly underestimated. On the other hand, the parametric Bayesian CDD approach outperformed the frequentist method, by producing more robust credible intervals and clear distinguishable posteriors distributions. It is noted that for both the frequentist and parametric Bayesian CDDs, the conditional marginals were assumed to follow a Beta distribution and the original marginals were uniformly distributed. This implies that the overall improvement was due to the uncertainty applied to the estimates of the functional of the dependence structure and not on the marginal estimation.
To conclude, our newly proposed parametric Bayesian CDD algorithm, can be used for the exploration and construction of gene interactions providing an insight into the architecture and functionality of genetic systems.

\section{Acknowledgements}
This work was supported by the ARC Centre of Excellence for Mathematical and Statistical Frontiers (Project ID: CE140100049). We would like to thank Dr Emily Wong from the Victor Chang Cardiac Research Institute for providing us with the data and for her help with the biological interpretation of the results.

%Bibliography
\bibliographystyle{plainnat} 
\bibliography{references}

\end{document}